\documentclass[11pt,preprint,prd,tightenlines]{revtex4}

\usepackage{graphicx}
\usepackage{amssymb}
\usepackage[dvips]{psfrag}

\begin{document}

   \title{Neutrino Oscillations for Dummies}

\author{Chris Waltham}
\affiliation{Department of Physics and Astronomy, University of British
Columbia, Vancouver BC, Canada V6T 1Z1}

\date{2003/03/28}

\maketitle

\section*{Abstract}

The reality of neutrino oscillations has not really sunk in yet. The
phenomenon presents us with purely quantum mechanical effects over
macroscopic time and distance scales (milliseconds and 1000s of km).
In order to help with the pedagogical difficulties this poses, I attempt
here to present the physics in words and pictures rather than math.
No disrespect is implied by the title; I am merely borrowing a term used
by a popular series of self-help books.

\section{Introduction}

Within the last two years, neutrino oscillations, first postulated in
1969, have become a reality.  The consequences for our view 
of
particle physics, astrophysics, and cosmology, are quite profound.
However, even for physicists, an understanding of the evidence for
oscillations and their mechanism can be elusive. This article aims to
present the current state of our knowledge of neutrino oscillations in as
simple a way as possible. It is a result of preparing for my talk at the
American Association of Physicists winter meeting of 2003, and of teaching 
the 4th-year particle physics course at UBC. A lot of
verbiage common to the neutrino oscillation industry will be explained.
For brevity and simplicity I will only look at what is currently the most
likely scenario for neutrino masses and mixings; other possibilities are
shamelessly ignored. The hope is that this treatment will be accessible to
physics teachers and other non-specialists. The reader is warned however
that this text is dense; most sentences contain some essential ingredient.
My bibliography aims at accessibility rather than rigour or completeness.
I present one mathematical equation.

\section{Quarks and Leptons}

The fundamental constituents of matter as we know it come in pairs - 
``doublets" - of elementary
particles which are very similar to each other except that their charges
differ by one fundamental unit. That unit is the magnitude of the charge
on the electron, which is the same as that on the proton. The first hint
that the universe is structured this way came when it was recognized that
atomic nuclei are formed of protons and neutrons, which are more-or-less
identical particles except that one has a charge of +1 and the other a
charge of zero. We now understand these particles in terms of their
elementary constituents, the quarks. The proton has two ``up"-quarks ($u$,
charge +2/3) and one ``down"-quark ($d$, charge -1/3); the neutron has two
downs and an up. In addition to these light, stable quarks, there are two
more doublets which have larger mass, and are unstable, eventually
decaying into ups and downs. These are called charm ($c$, +2/3) and
strange ($s$, -1/3), top ($t$, +2/3) and
bottom ($b$, -1/3). The names have no meaning, except to keep physicists
all talking
the same language. So we have three ``generations"  of quark ``flavours",
and this arrangement has deep significance for the structure of the
universe. Too deep to go into here.

\begin{center}
$(u,d), (c,s), (t,b)$
\end{center}

The link between the partners in the doublets is the weak-interaction, the
fundamental force which controls $\beta$-decay, and, as we shall see, 
solar
fusion reactions. The weak-interaction allows a $d$ to turn into a $u$
etc., where energy considerations permit. Significantly, it allows a free
neutron ($n$) to ($\beta$-)decay into a proton ($p$) and an electron
($e^-$, historically known as a $\beta$-ray).

\begin{equation}
n \rightarrow p + e^- + \bar\nu_e
\end{equation}

Or, in terms of the quarks (with the spectators in parentheses):

\begin{equation}
u(ud) \rightarrow d(ud) + e^- + \bar\nu_e
\end{equation}

We'll get to the $\nu_e$ in a second. Suffice it to say that without
neutron
decay there would be no free hydrogen in the universe, and we would
probably not be around to worry about neutrinos.

These flavour states ($u,d,c,s,b,t$) - we call them  ``eigenstates" - are
not kept rigidly
separate. The states in which these quarks propagate are called ``mass
eigenstates", which are each different mixtures of the flavour
eigenstates.
The mass
eigenstates propagate at different speeds and so the components get out of
phase with each other. A quark born as one flavour will soon start
looking like another.
An $s$-quark travelling through space can turn into a
$d$-quark. Likewise all the right-hand partners can mix between the
generations, and any particle made of 2nd or 3rd-generation quarks can
decay into a stable particle made of 1st-generation quarks.  By convention
we push all the mixing into the
right-hand partner; this is allowed because the weak-interaction allows
transformations between the left and right partners anyway.

So much for the structure of heavy particles (baryons), those which form
much of our
mass. What about the light particles (leptons), starting with the
electron? There are three charged leptons - electron, muon and tau - and
three associated very light chargeless neutrinos, clearly separated by one
unit of
charge.

\begin{center}
$(e^-,\nu_e), (\mu^-,\nu_\mu), (\tau^-, \nu_\tau)$.
\end{center}

Superficially this organization looks very much like the quarks.
However, for a long time the neutrinos were thought to be massless (as no
mass had been detected). In this case, no mixing is possible, as all
neutrinos will propagate at precisely the speed of light, and the mass
eigenstates can never get out of phase with each other. An
electron neutrino born will never change its  composition, as all parts 
will move at the same speed, and so the neutrino 
can never be detected as any other flavour.

As it turns out, this is not the case.

Mixing of flavours in the charged leptons would be easy to see. Muons
would decay into electrons and gamma-rays; this would be a fast
electromagnetic decay in contrast to the slow weak decay which actually
happens.

\begin{equation}
\mu^\pm \rightarrow e^\pm + \gamma
\end{equation}

\begin{equation}
\mu^+ \rightarrow e^+ + \bar\nu_\mu + \nu_e \ \ \mbox{or} \ \ 
\mu^- \rightarrow e^- + \nu_\mu + \bar\nu_e  
\end{equation}

The first decay violates lepton flavour and has never been seen.
The second decay tortuously conserves lepton number, and is the way all
muons seem to decay.

\section{Early Indications (1968-1992)}

By the 1960s our understanding  of the solar interior, and of low energy
nuclear physics, had reached such a stage that the Sun's neutrino output
could be predicted with some confidence. In broad terms, this picture
remains unchanged to this
day. For an excellent review of the subject in somewhat more mathematical
detail the reader is directed to an article by K. Nakamura in the Particle
Data Book\cite{nakamura}, which is available online.

\begin{figure}[h]
\includegraphics[width=10cm]{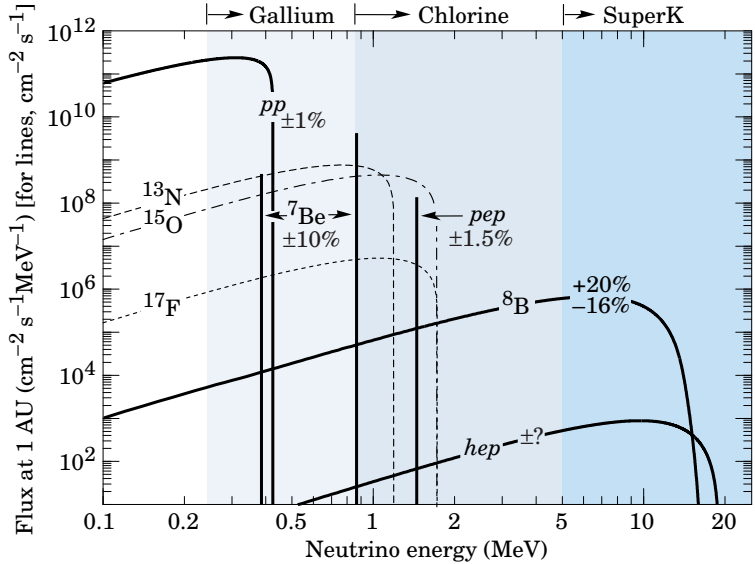}
\caption{The solar neutrino spectrum, showing the contribution of the
$^8$B and pp 
processes, and the thresholds for the three detection mechanisms. The 
figure
is reproduced by courtesy of the Particle Data Group.
\cite{pdg,bahcall}} 
\label{fig:spectrum} 
\end{figure}

The two most important reactions - both weak - yield the copious, low
energy
``pp" neutrinos, and the more scarce high energy ``$^8$B" neutrinos
(see fig.\ref{fig:spectrum}).

\begin{equation}
p + p \rightarrow \mbox{D} + e^+ + \nu_e \ \ \ \ E_\nu = 0-0.420 \mbox{ 
MeV}
\label{eqn:cc}
\end{equation}

\begin{equation}
^8\mbox{B} \rightarrow 2^4\mbox{He} + e^+ + \nu_e \ \ \ \ E_\nu = 0-14.6
\mbox{ 
MeV}
\end{equation}

The ``D" is a deuteron, heavy hydrogen $^2$H.
The pp-neutrinos are very hard to see, even by neutrino standards, and can
only be detected by radio-chemical means (see below). However their flux
(about 10$^{15}$~m$^{-2}$s$^{-1}$ at the Earth) is 10,000 times that of
the
$^8$B-neutrinos. (An MeV is a million electron volts, a typical energy
release for a single nuclear reaction.)

Now the experimental nuclear physicists took up the challenge to detect
these elusive particles.

\subsection{Neutrino Detection}

The chargeless neutrino cannot be ``seen" unless it is (a) absorbed 
on a nucleon and produces a charged lepton, or (b)
scatters on an electron. In both cases the reaction products may be seen 
- always
through their electromagnetic properties. There are two
fundamental ways of detecting a neutrino; either by scattering off a
nucleon,
or an electron.

\subsubsection{Nucleons}

Electron-type neutrinos can scatter off a neutron to produce a proton, and
the other
half of the neutrino's doublet, the electron. This is a charge-swapping
``quasi-elastic" reaction.

\begin{equation}
\nu_e + n \rightarrow e^- + p
\end{equation}

This reaction is exothermic, as the chemists would have it, and so would
be accessible to all solar
neutrinos. Unfortunately, free neutrons don't live much longer than a
quarter of an hour, and so a macroscopic detector is hard to build. The
reaction doesn't work on protons (hydrogen would be a convenient
detector) because the charges cannot be made to match (it will for {\it
anti-}neutrinos, which can make positrons - $e^+$).

The next best thing is to use neutrons in a nucleus. This is not easy as
protons are usually bound less tightly in a nucleus than are neutrons, and
so their effective mass increases and this raises the energy threshold for
the reaction, in most cases out of reach of solar neutrinos. In addition, 
the outgoing electron is usually too low in energy to see,
and so
one has to rely on a detectable product nucleus. This means it has to be
physically removable from the detector and radioactive with a
convenient-enough half-life (preferably days) to accumulate and to observe
after removal. In fact the reaction tends to make a nucleus which is too
proton-rich, and these do tend to decay by the slow process of
atomic-electron
capture, which produces detectable X-rays. The basic decay process is the
reverse of neutrino absorption and yields back the original nucleus:

\begin{equation}
e^- + p \rightarrow \nu_e + n
\end{equation}

In addition, the weakness of the weak interaction means we need 100s
or 1000s of tonnes of target material to get a good signal, so the
substance has to be cheap, safe and purifiable of trace
radioactivity. Three nuclei have been used as
detectors to date. The first was chlorine-37, in the form of borrowed
cleaning fluid, which has been used by Ray Davis\cite{davis} since the
late '60s:

\begin{equation}
\nu_e + ^{37}\mbox{Cl} \rightarrow e^- + ^{37}\mbox{Ar}
\end{equation}

This method of detection is sensitive to most of the $^8$B spectrum and
neutrinos 
from some
intermediate reactions. The radioactive argon, a noble gas,  can be
bubbled out of the
chlorine using helium. It decays by the capture of an atomic electron, a 
process which kicks out further atomic (``Auger") electrons which can be 
detected and counted. 

\begin{equation}
e^- + ^{37}\mbox{Ar} \rightarrow \nu_e + 
^{37}\mbox{Cl} + \mbox{Auger electrons}
\end{equation}

This was the first
successful detection method for solar neutrinos, and as a result, Davis
shared the 2002 Nobel Prize in Physics\cite{nobel}.

The only detector successfully used to date which can see the pp-neutrinos
is gallium-71. The first results were recorded by the Soviet-American
Gallium Experiment SAGE\cite{sage}, and the Italo-German Gallex
\cite{gallex}, in 1990.

\begin{equation}
\nu_e + ^{71}\mbox{Ga} \rightarrow e^- + ^{71}\mbox{Ge}
\end{equation}

It works like the chlorine reaction except that gallium is converted to
radioactive germanium eve. However, you
can't bubble germanium out of liquid gallium without first converting it
to a gas. This has been done, but it requires more chemistry than you can
shake a stick at. Wags call the gallium-germanium-gallium sequence the
``Alsace-Lorraine" process, but don't tell this to an undergraduate class
unless you want rows of blank faces.

The third useful nucleus is deuterium,  but in a
different way. The proton
and neutron are bound in deuterium so lightly that the electron {\it is}
energetic enough to be visible. This is fortunate, because the stable
protons are not:

\begin{equation}
\nu_e + \mbox{D} \rightarrow p + p  + e^-
\end{equation}

The most convenient form of deuterium is heavy water (not cheap but at
least available\cite{d2o}), and the fast electron emits Cherenkov
radiation which
can be
readily detected. Cherenkov radiation is the result of a charged particle
travelling at faster than the local speed of light (in water = c/1.33)
which emits the electromagnetic equivalent of a supersonic boom, a conical
pattern of blue and UV photons. The first such heavy water Cherenkov
detector is the Sudbury Neutrino Observatory (SNO)\cite{SNO}, and it
started
operating in
1999, a decade after the first light water Cherenkov detector, Kamiokande
(see next
section).

\subsubsection{Electrons}

Conceptually the simplest way to detect neutrinos is via elastic
scattering off electrons, which are plentiful in any material. One simply
has to choose a cheap, safe, purifiable, transparent medium so the
Cherenkov light from neutrinos can be distinguished from the inevitable 
trace radioactivity. Water works well. The masters of this technique are
the Japanese Kamiokande collaboration (and its successor,
SuperKamiokande). 
Kamiokande announced the first real-time, directional detection of solar
neutrinos in 1989\cite{kamII}. It was for initiating this series of 
experiments
that Masatoshi Koshiba shared the 2002 Nobel Prize in Physics\cite{nobel}. 
The
reaction is:

\begin{equation}
\nu_e + e^- \rightarrow \nu_e + e^-
\end{equation}

Undergraduates have complained to me that this is not a real reaction
because the left hand side is the same as the right (and get even more 
confused if I occasionally leave the $+$ signs out). They have obviously
been made to sit through too many chemistry lectures. The left-hand
electron is stationary, while the right-hand one is recoiling from the
neutrino and is moving close to the speed of
light, which makes it visible. 
The directionality comes from the fact that the recoil electron tends to
travel
in line with the original neutrino.
This method of detection is sensitive to
the upper end of the
$^8$B spectrum. 

There's a slight but significant complication here. Electron scattering
also works for muon and tau neutrinos, e.g.:

\begin{equation}
\nu_\mu + e^- \rightarrow \nu_\mu + e^-
\end{equation}

The cross-section (i.e. sensitivity) of this reaction is only 15\% of that
for electron neutrinos.

\subsubsection{The Data}

By 1990, the chlorine, light water, and gallium experiments (in that
order), had all reported
seeing only a small fraction of the expected signal. The numbers are 
given below. Note that all these
experiments have to be done deep underground, to get away from cosmic rays
which would easily swamp the tiny neutrino signal on the surface. 

\begin{itemize}

\item Gallium (0.2 MeV threshold) - 55\% 

\item Chlorine (0.8 MeV threshold) - 34\% 

\item Light Water (9 MeV threshold, later 5 MeV) - 48\%

\end{itemize}

There were 10-20\% errors in these numbers due to experimental and
theoretical uncertainties, but it was clear that none of these numbers was
remotely consistent with 100\%.
By this time, the solar astrophysicists were getting very confident of
their flux predictions, and so these deficits became known as the ``Solar
Neutrino Problem". When physicists use the word ``problem", they mean
``funding opportunity".

\section{Pontecorvo's Idea}

All the above detection reactions are sensitive only to $\nu_e$ (with
the one
 exception noted). In 1969 Bruno Pontecorvo reasoned that if
neutrinos had small and different masses, plus flavour mixing, then 
$\nu_e$s born in the Sun might reach Earth as, say $\nu_\mu$ ($\nu_\tau$s
were
not known then) and be undetectable\cite{bp}. This might be the answer to
the 
emerging Solar Neutrino Problem.

For simplicity, consider two  neutrino
species. Neutrinos
are born and detected via the weak interaction as ``flavour eigenstates",
e.g. $e$ and $\mu$ . However they
propagate as ``mass eigenstates" which have a distinct velocity, labelled,
for example, ``1"
 and ``2" 
 
If flavour is not rigorously conserved (there is no particular reason why 
it should be, except that's the way the charged leptons seem to behave),
and if the masses $m_1$ and $m_2$ are
slightly different, these two pairs of states may not be one and the same,
but may be mixed:
                                 
\begin{equation}                
\nu_1 =  \nu_e \cos\theta + \nu_\mu\sin\theta \ \ \ \ \
\nu_2 =  \nu_e \sin\theta + \nu_\mu\cos\theta
\end{equation}

Our use of sines and cosines of an angle ensures that the mixing produces
neither more or less neutrinos than we started with. Its called a
``unitary" transformation because the particle number depends on the
square of the amplitude terms shown ($\cos^2 + \sin^2$ is always 1 - Math 
10).
We will define the phrase ``slightly different" later. Consider a neutrino
which was
created as an electron neutrino. The probability that it will be detected
as an electron neutrino a
distance $L$ away is (after a bit of math):
 
\begin{equation}
P_{ee} = P_{\mu\mu} = 1 - \sin^22\theta \sin^2 k\Delta m^2L/E
\label{eqn:pee}
\end{equation}

\begin{figure}[h] 
\includegraphics[width=8cm,angle=-90]{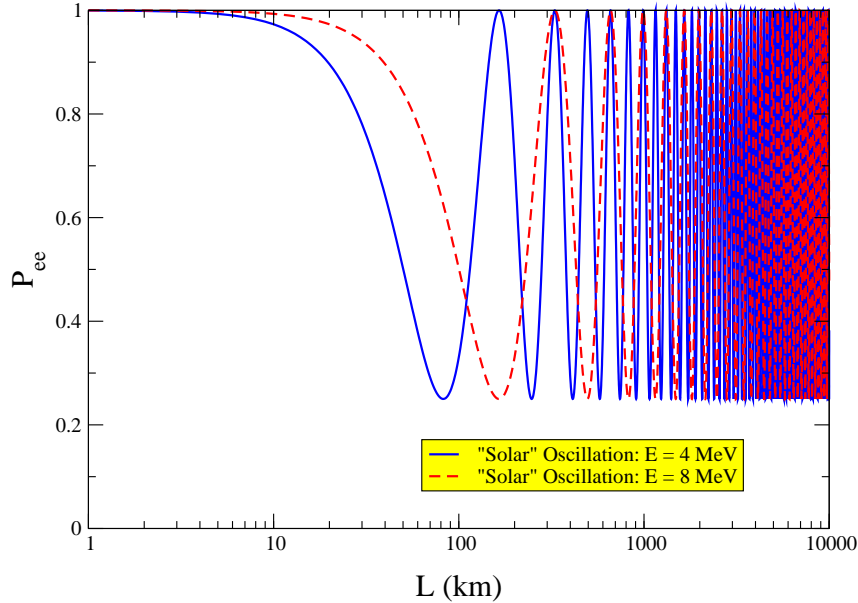}
\caption{The probability $P_{ee}$ is plotted against the distance between 
the
source and detection for two monoenergetic electron neutrinos, at 4 MeV
(typical
detected reactor
neutrino energy) and at 8 MeV (typical detected solar neutrino energy).
Both have the currently accepted solar oscillation parameters of
$\Delta m^2_S = 6\times 10^{-5}$ and $\theta = \pi/6$.
It is clear that a reactor experiment, with a neutrino spectrum several
MeV wide, will see no effect below 10~km,
but at further than 100~km will see about 60\% of the unoscillated flux.
}
\label{fig:osc} \end{figure}

Numerically, the constant                           
$k = 1.27$ if $\Delta m^2 = m_2^2 - m_1^2$ is measured in eV$^2$ (strictly
eV/$c^2$ all squared, from $E=mc^2$ -  but we like to think $c$ = 1), the
energy $E$ in MeV and the distance
from
the source $L$ in m. It will equally work for GeV and km.
These are classical ``vacuum oscillations", and they are plotted for the
parameters we now associate with solar neutrinos in fig.\ref{fig:osc}.

With {\it three} species, the mass and flavour basis states are linked via
a
$3\times3$
matrix
with two mass splittings, three angles like $\theta$ and an extra one
which makes life different for neutrinos and antineutrinos (a
``CP-violating phase").

\begin{figure}
\includegraphics[width=7cm]{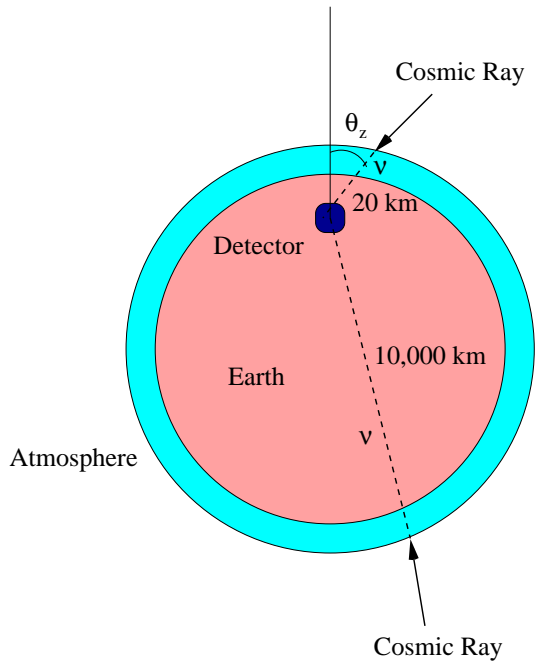}
\caption{The production mechanism and geometry for the detection of
atmospheric neutrinos.}
\label{fig:atmospheric}
\end{figure}

\begin{figure}[h] 
\includegraphics[width=8cm,angle=-90]{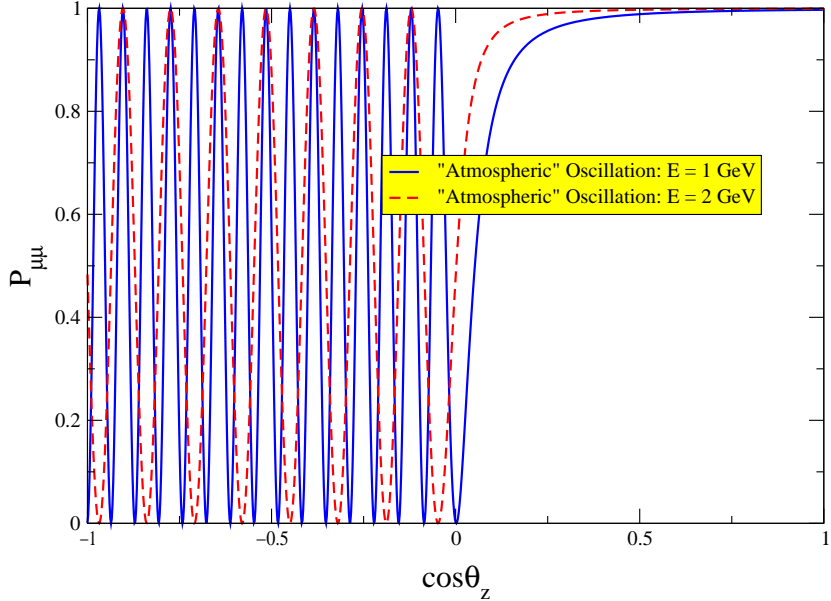}
\caption{The probability $P_{\mu\mu}$ is plotted against the cosine
of the zenith angle $\theta$ for the case of an atmospheric neutrino
experiment. Neutrinos coming from directly above have $\cos\theta = +1$,
those from the nadir have $\cos\theta = -1$.
The energies plotted  are  1 and 2 GeV (typical values in this case)
with the accepted atmospheric  oscillation parameters of
$\Delta m^2_A = 3\times 10^{-3}$ and $\theta_A = \pi/4$.
It is easy to see that with an atmospheric neutrino spectrum a few GeV
wide, there will be no effect on $\nu_\mu$ flux above the horizontal,
but only 1/2 of the expected $\nu_\mu$ flux will be seen below the
horizontal.
}
\label{fig:zenith} \end{figure}

\section{The Hard Evidence, 1992-2002}

There are now two widely recognized pieces of evidence for neutrino
oscillations:
 
\begin{itemize}
\item the atmospheric muon-neutrino deficit 
\item  the solar electron-neutrino deficit 
\end{itemize}
 
The first was formally announced by the Kamiokande collaboration in 1992
\cite{kamiokande}. It was declared to be proof of non-zero neutrino 
mass (consistent with oscillations) in
1998 by that collaboration's successor, SuperKamiokande \cite{SK}.
Evidence for the latter had been building steadily over 30 years, but
proof came in 2001/2 by the Sudbury Neutrino Observatory
collaboration\cite{sno2002},
who showed that solar astrophysics could not be to blame for the deficit,
and that the ``missing" neutrinos were arriving at the Earth as other
flavour states. The particular oscillation mechanism suggested by the
solar experiments was confirmed in December 2002 by the 
KamLAND reactor-neutrino detector, a fact which removed any lingering
worries about uncertainties due to solar astrophysics.

We will call the neutrino parameters revealed by atmospheric neutrinos
$\Delta m^2_A$ and $\theta_A$, and those revealed by solar (and reactor) 
neutrinos
$\Delta m^2_S$ and $\theta_S$. The remaining angle (``$\theta_{13}"$) is
now very much sought after. The CP-violating phase is even
hotter
property but it will be very hard and expensive to find\cite{jhf}.

\subsection{Atmospheric Neutrinos}

Atmospheric neutrinos are made by pion and kaon decays resulting
from cosmic-ray
interactions in the upper atmosphere. The numerology of these decays leads
one to expect two $\nu_\mu$s for each $\nu_e$ (it is very hard
experimentally to distinguish between neutrinos and antineutrinos,
($\bar\nu$). The following reaction sequence is typical:

\begin{equation} 
p+ ^{14}\mbox{N} \rightarrow \pi^+ + X \ \ \ \ \ 
\pi^+ \rightarrow \mu^+ \nu_\mu \ \ \ \ \ 
 \mu^+ \rightarrow e^+  \nu_e  \bar\nu_\mu 
\end{equation}

Here $X$ means nuclear fragments.
However, the observed $\mu/e$ ratio
is more
like unity on average, and is strongly dependent on zenith angle i.e. the
distance
the neutrino has travelled since birth (fig.\ref{fig:atmospheric}). It
seems that
the further a
$\nu_\mu$
travels in the Earth, the less chance it has of being detected, even
though the probability of absorption is tiny. In fact the mean free path
of an atmospheric neutrino in rock is about $10^{12}$~m, so only a tiny
fraction interact in the Earth with its diameter of only
$10^7$~m. Therefore they must be disappearing by some other means.

 The measured
flux of $\nu_\mu$s is about half the expected value, while that for the
$\nu_e$s
is
about right. There are strong indications from SuperKamiokande (SK) that
the missing $\nu_\mu$s are showing up
as $\nu_\tau$s (which are very hard to see). We will assume this is the
case.
 
Detailed analysis in terms of path length through the Earth yields an L/E
dependence as expected from the eq.\ref{eqn:pee}, with
parameters:
 
\begin{equation}
\Delta m^2_A \approx 3\times10^{-3} \mbox{ eV}^2 \ \ \ \ \  \theta_A 
\approx
\pi/4
\end{equation}

Atmospheric neutrinos are typically a few GeV in energy, so 
$\Delta m^2_A$
is more-or-less fixed by the geometry of the earth, otherwise the
effect would be unobservable. The data sample is
divided up into two distinct angular regimes: the neutrinos are coming
either from above, with path
lengths of 10s of km, or from below, with path lengths of 1000s of
km (fig.\ref{fig:atmospheric}). Simple solid-angle considerations tell us
that not many neutrinos come
from around the horizontal, with
path lengths of 100s of km. One reason why this evidence is so solid,
despite the complexity of the particle interactions in the atmosphere, is
that we see no oscillation effects
above the horizontal, and plenty below it. The only possible complications
are those of Earth's geometry and magnetic
field (which distorts the isotropy of the cosmic rays), but these are now
well understood.
 
Hence, oscillations like this will be visible for 1000 km pathlengths and
1 GeV neutrinos if $\Delta m^2_A$ is around $10^{-3}$~eV$^2$, and the
mixing angle is big, as
can readily be seen from
eq.\ref{eqn:pee} and fig.\ref{fig:zenith}. This is precisely our
conclusion.

The SK collaboration announced this result as the discovery of neutrino
oscillations in 1998. The big surprise is the largeness of the mixing
angle, which may be maximal
(45$^\circ$). The mixings previously observed in the quark sector are very
small
(the biggest is 13$^\circ$).
So, two of the mass eigenstates (actually 1 and 3) are likely an equal mix
of $\nu_\mu$ and $\nu_\tau$, with very little $\nu_e$ in $\nu_3$. 
 
The size of the neutrino source (a few km layer in the upper
atmosphere) is small compared to flight distances (20 km -13,000 km), and
$\nu_\mu$ and $\nu_\tau$ interact identically in the
earth (i.e. no matter effects) so it is useful to think in terms of
vacuum
oscillations. Things are rather different in the solar case...

\subsection{Solar Neutrinos}

The final piece of evidence for solar neutrino oscillations came in 2002
when the SNO collaboration reported two numbers, one for the flux of
electron neutrinos measured by reaction \ref{eqn:cc}, and one for the
total flux of all neutrinos flavours measured by a reaction unique to
deuterium, which is totally blind to neutrino flavour ($x=e,\mu,\tau$):

\begin{equation}
\nu_x + \mbox{D} \rightarrow p + n  + \nu_x
\end{equation}

Here the neutron is detected by capture on deuterium, which produces a
visible gamma ray. 

The analysis of this reaction in SNO's data yields precisely the flux of
neutrinos expected from standard solar astrophysics. However, comparison
with events of the type shown in eq.\ref{eqn:cc} show that only 34\% of
these neutrinos still have electron flavour by the time they reach us.
The remaining 66\% are in all likelihood an equal
mix of $\nu_\mu$ and $\nu_\tau$, if the atmospheric interpretation is
correct.
It should also be noted that there is some evidence that the electron
fraction increases at night, when the neutrinos have passed through 1000s
of km of the Earth to get to the detector. 

In retrospect, the relatively large signal in the light water detectors
was a good clue: some of this is due to $\nu_\mu$ and $\nu_\tau$.
Once this is taken into account, the exact fraction observed varies
only slightly
with energy. The SNO and SK
observatories see no distortion; only the ultra-low threshold gallium
experiments see a slightly larger fraction.
This 
is not what you would expect from vacuum
oscillations, which have a strong energy dependence. However, another
process is at play,
 known as the ``Large Mixing Angle
MSW" or
LMA
solution which relies on the behaviour of neutrinos in the dense interior
of the Sun.

\subsubsection{Neutrinos in Dense Matter}

\begin{figure}
\includegraphics[width=10cm]{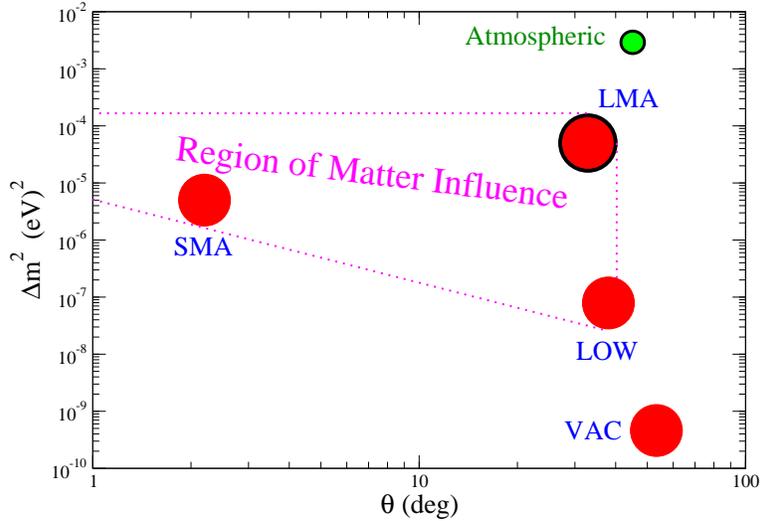}
\caption{Various solutions to the solar neutrino which were popular prior 
to SNO's 2002 papers, plotted in $\Delta m^2-\theta$ space. The vacuum 
(VAC), ``low" (LOW) and small mixing angle (SMA) solutions are now thought 
much less likely 
than
the large mixing angle (LMA) solution. The region where matter effects 
come into play is shown, as is the solution to the atmospheric neutrino 
problem, for comparison.} \label{fig:msw}
\end{figure}

The ``MSW" effect refers to three theorists Mikheyev, Smirnov and
Wolfenstein who first explained how neutrino mixing would be affected in
the presence of dense matter - the solar core for example, or perhaps the
interior of the Earth. 
Because ordinary matter contains electrons, and not $\mu$ or $\tau$, 
$\nu_e$ behave in a subtly different manner in matter than do
$\nu_\mu$ \ or $\nu_\tau$. This distorts the masses and flavour components 
of the mass eigenstates when neutrinos move between a vacuum and dense 
matter. The region of parameter space in which these effects are important 
is shown in fig.\ref{fig:msw}. It is fairly narrowly defined in terms of
$\Delta m^2$. If $\Delta m^2$ is too big, then matter effects become 
negligibly small - the critical value is fixed by fundamental physics and 
the density of the solar core to be around 10$^{-4}$~eV$^2$. If $\Delta 
m^2$ is too small, then the vacuum oscillation length becomes much bigger 
than the solar core and so this region of dense matter starts to look too 
small, from the neutrino's perspective, to have an effect.

On fig.\ref{fig:msw} is marked what looked like plausible solutions to the 
solar neutrino problem in 1987, when the idea of matter effects arose.
The small mixing angle ``SMA" solution was the theoretical favourite, as 
small mixing angles were what we were used to from the behaviour quarks.
However, this and the vacuum solution (``VAC") produced strong spectral 
distortions which we just don't see. The LMA solution reduces the $\nu_e$
flux evenly across the spectrum, 
and 
by 2002 it had emerged as the likely truth.

A detailed analysis yields the following oscillation
result:

\begin{equation}
\Delta m^2_S \approx 6\times10^{-5} \mbox{eV}^2 \ \ \ \ \  \theta_S
\approx \pi/6
\end{equation}

You can now see that the vacuum oscillation length is a few 100 km. This
is small compared with the 35,000 km radius of the solar core where these
neutrinos are born. Thus
vacuum oscillation effects are totally washed out. So, for understanding 
solar neutrinos: THINK MASS
EIGENSTATES!

\begin{figure}
\includegraphics[width=8cm]{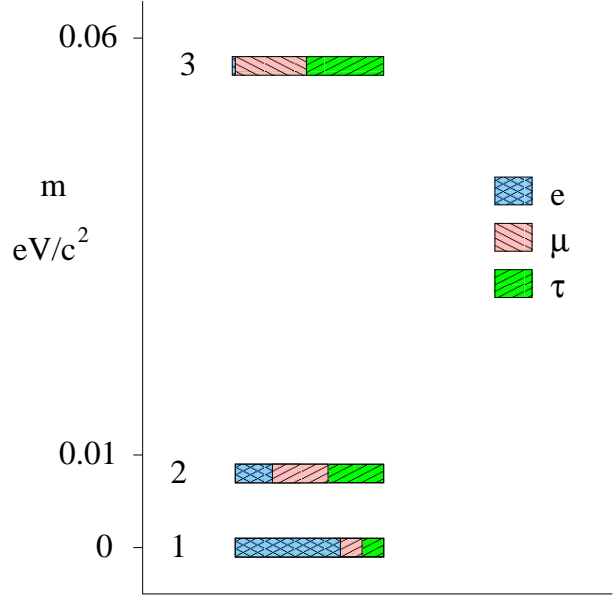}
\caption{
The flavour composition of mass eigenstates assuming the LMA solution 
for solar neutrino oscillation.  
The electron component of $\nu_3$ is not confirmed
and may be very small. The mass scale assumes that $m_1$ is zero, which
may not be the case. The smaller mass difference may be 2-3 rather than
1-2.
}
\label{fig:vacuum} 
\end{figure}

The flavour composition of mass eigenstates in the LMA scenario is shown 
in 
fig.\ref{fig:vacuum}. The simplest mathematical expression for the 
composition of neutrino
mass eigenstates in a vacuum, consistent with the data, is as 
follows\cite{jelley}:

\begin{eqnarray}
\nu_1 & \approx & \sqrt{\frac{3}{4}}\nu_e - \sqrt{\frac{1}{8}}(\nu_\mu - 
\nu_\tau) \\
\nu_2 & \approx & \sqrt{\frac{1}{4}}\nu_e + \sqrt{\frac{3}{8}}(\nu_\mu - 
\nu_\tau) \\
\nu_3 & \approx & \sqrt{\frac{1}{2}}(\nu_\mu + \nu_\tau)
\end{eqnarray}

\begin{figure}
\includegraphics[width=15cm]{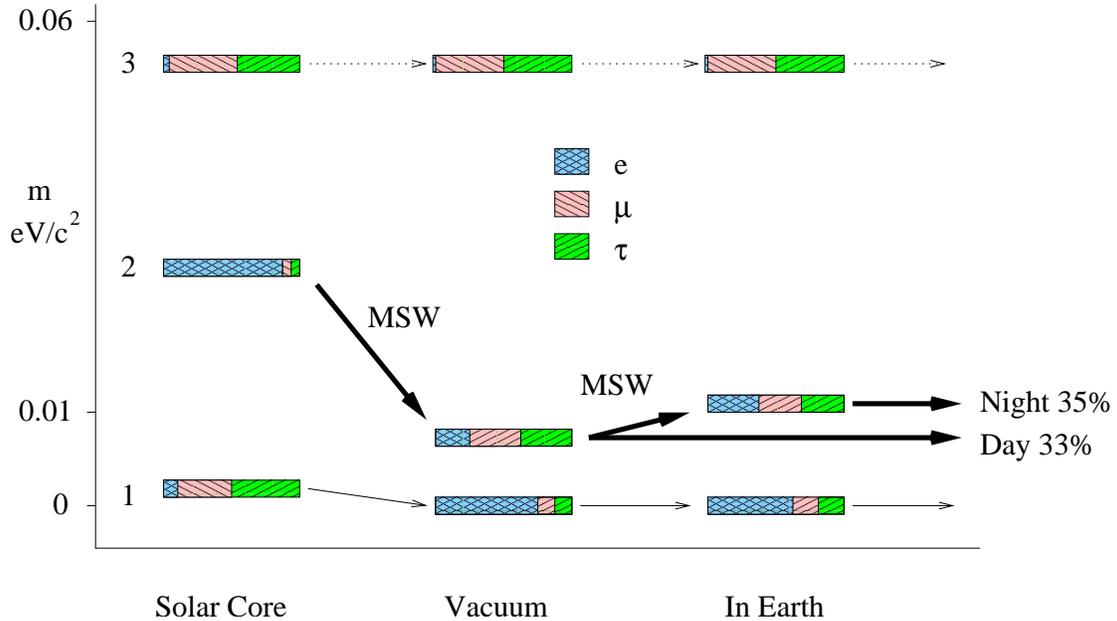}
\caption{A cartoon of the LMA solution for solar neutrino oscillation.  
Here is the big picture of solar neutrino physics - including what we know
about atmospheric neutrinos - in terms of mass eigenstates and their
flavour components.
Representing the flavour components as fixed fractions is only really
useful here because the oscillations are washed out due to the source
size.  I assume non-inverted hierarchy and $m_1 = 0$. At the moment there 
is no evidence for the electron component in $\nu_3$; shown here is
roughly the
maximum possible value. The electron component of $\nu_1$ and the 
non-electron component of $\nu_2$ are likely very small. I have 
also exaggerated 
the regeneration of $\nu_e$ in the earth; the effect may be very tiny. 
}
\label{fig:big} 
\end{figure}

Whether this scenario is approximately or exactly correct is one of the 
biggest questions in neutrino physics at present. Evidence, or lack 
thereof, for a $\nu_e$ component in $\nu_3$ is crucial, as noted above.
To understand how we come to this result one has to backtrack
the neutrinos from a vacuum, into the heart of the Sun. As the rising
density starts to single out the $\nu_e$ as ``special", nearly all the 
electron flavour is piled into $\nu_2$, whose effective mass rises because 
of the preferential interaction between $\nu_e$ and electrons.
This is shown in fig.\ref{fig:big}. The effect is somewhat energy 
dependent, but above a neutrino energy of 5~MeV, the $\nu_e$ component of 
$\nu_2$ has risen 
from a vacuum value of 25\% to $>99$\%. 
For a lightly mathematical account of how this arises, see 
ref.\cite{kayser}.

Hence when a  $\nu_e$ is born in a fusion or decay
reaction, it becomes mostly $\nu_2$, with a bit of $\nu_1$, and possibly a
tiny bit of $\nu_3$. These mass states then proceed out of the Sun and on
to the Earth. In doing so, $\nu_2$ ends up a fairly equal mix of all
flavours, and that is basically what we measure (along with possibly 
tiny 
contributions 
from  $\nu_1$ and  $\nu_3$). 

There are hints in SNO's data that the $\nu_e$ fraction in observed solar 
neutrinos is 
slightly higher at night. The LMA scenario can easily explain this by 
allowing the $\nu_2$ to become a little more electron-like after taking a 
good long run through the dense Earth at night.

A word of caution regarding fig.\ref{fig:big}. The electron component of 
$\nu_3$ is not confirmed
and may be very small. The mass scale assumes that $m_1$ is zero, which
may not be the case.
In addition, the smaller mass difference may be 2-3 rather than
1-2; we cannot tell as yet. In any case, these mixings are fundamental, 
and the next challenge is to figure out why they are this way.

The solar results were confirmed in December 2002 by the Japanese
ultralong-baseline reactor experiment
KamLAND\cite{kamland}. Here we have
a point source, and it is
therefore OK to revert to thinking of vacuum oscillations. 
A look at fig.\ref{fig:osc} shows why all previous reactor experiments
with shorter baselines saw no effect, while KamLAND sees 60\% of the
unoscillated flux at 180~km. Note that the difficulty of these experiments 
increases rapidly with distance, because even without oscillations, the
inverse-square law is in effect, and one rapidly ends up with no signal at
all.

\section{Big Questions}

\subsection{ Why don't charged electrons and muons oscillate? }

  Electrons, muons (and taus) are charged. The things that we know
oscillate ($K^0$s, neutrinos) are neutrals. We know two things about
neutrals: we can deduce that they
  were made (by observing other reaction products), and we can deduce that
they died (by decay or absorption products). What goes on in between is
anyone's guess (i.e.
  quantum mechanics). You can figure out that you just made an
electron neutrino, and somewhere down the beam pipe figure out that a
muon
  neutrino just interacted and died. 
The electron and muon neutrino do not have well defined masses (they are
not mass
eigenstates), but the mass
  splittings are tiny, so whatever oscillation occurs, we never have to
worry about mismatches in measured energies or momenta\cite{lowe}. 
In between
life and death, however, they
  travel as mass eigenstates ($\nu_{1,2,3}$ )
   
  Charged particles are different, we can track them every mm of the way
through a drift chamber. Like Schr\"odinger's cat, they interact too much
with the environment to
  be in an ill-defined state. In addition the mass differences are huge
(at least 100 MeV, a billion times the biggest neutrino mass difference),
so if you watch a muon suddenly become an electron for a few metres and
then revert to a muon,
  there would be major accounting difficulties on the energy/momentum
front. For muons and electrons, the mass eigenstates and flavour
eigenstates are (have to be?) one and the same
  thing.

\subsection{Why are the mixings so large?}

We don't know.  A clue will come when we measure
the electron component in $\nu_3$ at Japanese Hadron Facility in the next
decade\cite{jhf}. This angle may be merely small, or tiny enough to be a
clue to some new physics. This will also tell us whether the 
$\nu_\mu-\nu_\tau$
mixing is actually, or merely approximately, maximal.
This will tell us whether the angles are somehow randomly chosen or in
some way ``special".

\subsection{If neutrinos exit the Sun without interaction, how does the 
density of the solar core affect them at all?}

This one is the hardest to explain to the man at the bus stop. Its all to 
do with amplitudes and probabilities. 
Probabilities (of interaction, say) tend to be the squares of quantum 
mechanical amplitudes. A solar neutrino can pass through a light year of 
lead with only a small chance of interaction, so the probability here is 
tiny. But the amplitude (its square root) is not that small. The way the 
electron density in the solar core affects neutrinos depends, however, on 
amplitudes. Hence is it possible to alter the neutrino states while the 
probability of interaction is negligibly small. The characteristic 
distance required for this skewing of neutrino states is only about 100~km 
in the core of the Sun, and a few 1000s of km in the 
Earth\cite{bahcall_book}. 
This is one 
reason why the day-night asymmetry, if it can be measured at all, is very 
small. The Sun never dips too far below the horizon even at the most 
southerly solar neutrino detector (SuperKamiokande, 36$^\circ$ N).

\subsection{What are the absolute masses of the neutrinos?}

\begin{figure}
\includegraphics[width=8cm, angle=-90]{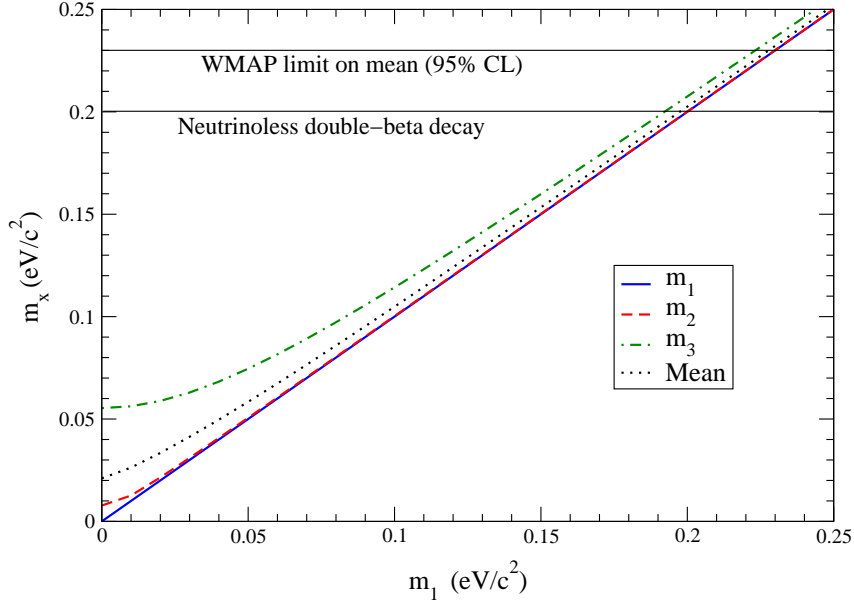}
\caption{Merging neutrino oscillation evidence with absolute mass limits
from measurements of the cosmic microwave background and double-beta 
decay. The masses $m_{1,2,3}$ are plotted
for varying values of $m_1$, from zero up to the absolute mass limit.} 
\label{fig:dm2} 
\end{figure}

We're closing in on this question. We have the splittings, and they are
plotted as a function of $m_1$ in fig.\ref{fig:dm2}. The minimum value
$m_3$ can have is about 0.06 eV/c$^2$. Observation of ripples
in the cosmic microwave background\cite{WMAP} and 
double-beta decay experiments\cite{vogel} set the maximum to be about 0.2
eV/c$^2$.
Another decade of hard work should close this gap.

\subsection{What is the mass of an electron (or mu or tau) neutrino?}

The $\nu_e$ doesn't have a well-defined mass; it is a thorough mix of two
(or three) neutrinos states which do have well-defined, but different
masses. If evidence of a neutrino mass is seen in $\beta$-decay spectra, 
it will be an averaged value (depending on the mixing angles) of these 
mass eigenstates.

\subsection{Will we be able to see the decay $\mu\rightarrow e\gamma$? }

\begin{figure}
\includegraphics[width=10cm]{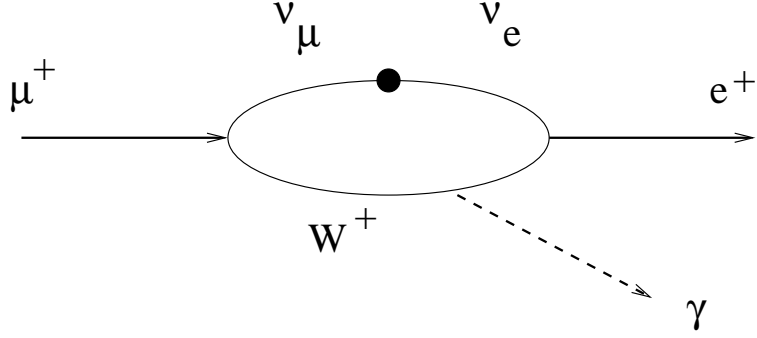}
\caption{Feynman diagram for $\mu\rightarrow e\gamma$. Imagine time
running left to right. The heavy $W^+$ can link a muon with a muon
neutrino, or an electron with an electron neutrino, and also - because it 
is charged - shake off a gamma ray. Hence if the neutrinos mix, so must
the muon and electron. However, the big mass-energy of the $W^+$ can only 
be borrowed for a brief $10^{-26}$s, too short for the leisurely neutrino 
oscillation to take place.} 
\label{fig:meg} \end{figure}

No. 
By the tenets of the Uncertainty Principle one can, for a very short
period of time, turn a  muon into a heavy $W$ and a $\nu_\mu$
(fig.\ref{fig:meg}). This
$\nu_\mu$ can in principle oscillate into a $\nu_e$, which can coalesce
with the $W$ into an electron. The $\gamma$-ray carries away the excess
energy and momentum and everyone is happy. However, Heisenberg tells us we
can borrow 80 GeV for a $W$ for only 10$^{-26}$s or so. But we already
know that it takes 1000s of  km for a 100 MeV (the muon's mass) neutrino 
to
oscillate. That's many milliseconds (an eternity) at the speed of light, 
so its 
never
going to happen.

\section{A Musical Analogy}

\begin{figure}
\includegraphics[width=15cm]{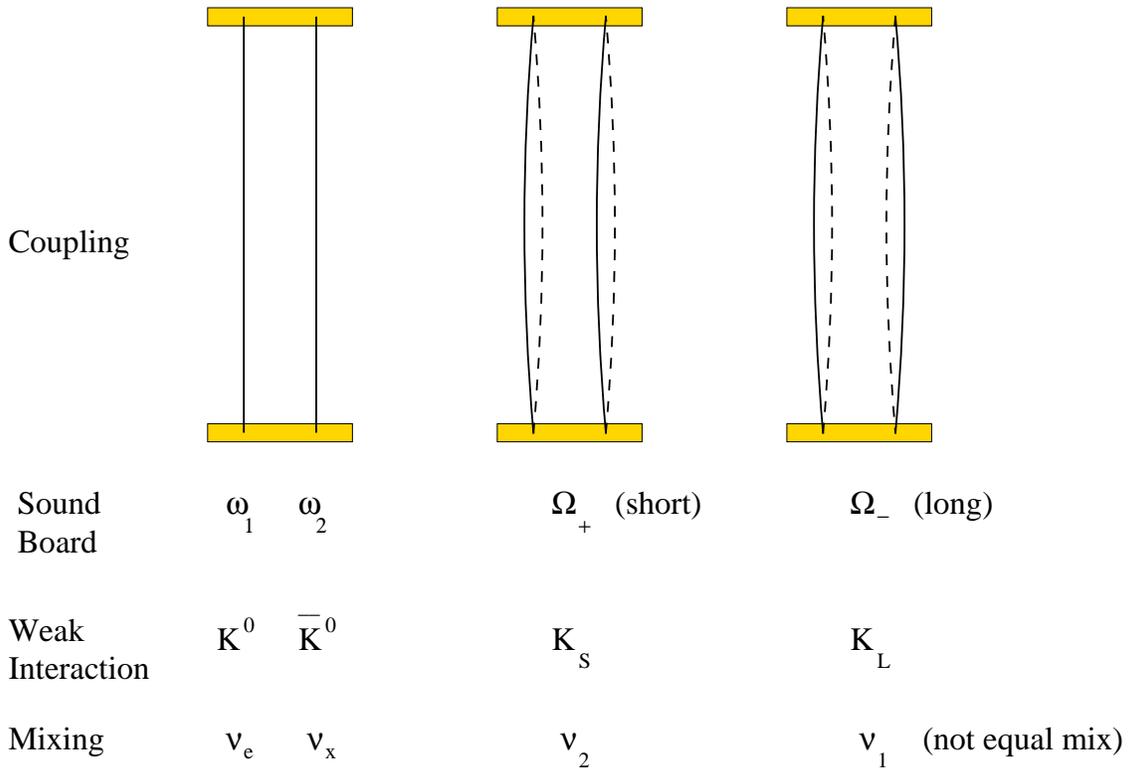}
\caption{The analogy between piano strings, neutral kaons, and vacuum 
oscillations. That for neutral kaons is rather exquisite, but for 
neutrinos one has to remember that the mixture of $e$ and $x=\mu-\tau$
in $\nu_{1,2}$ is not equal but in the ratio 
$\frac{1}{4}$::$\frac{3}{4}$.}
\label{fig:music} \end{figure}

Consider two adjacent piano strings, tuned to very nearly the same pitch,
($\omega_1$, $\omega_2$) attached to the sound board. Pluck or strike one 
string, and soon both strings will be oscillating with a mixture of the 
two 
modes, one with the strings in phase, one with them in antiphase. 
The frequencies of these modes are $\Omega_+$ and $\Omega_-$, as shown in 
fig.\ref{fig:music}. The secret of the piano sound is that the $\Omega_+$ 
mode couples very strongly to the sound board and makes a loud but rapidly 
decaying sound. The $\Omega_-$
mode couples only weakly to the sound board (the motion of one string
negates the other) and produces a soft but sustained 
sound\cite{weinreich}.

This is a useful approximate analogy for neutrino vacuum oscillations. 
Each string can be considered a flavour eigenstate, but the time 
development goes on in terms of the $\Omega_\pm$ modes. Averaged over 
time, half the energy appears in each string, regardless of which string 
is struck.
However, the neutrino mixing is not 50-50 (at least not for solar 
neutrinos), and neutrinos do not decay, as does a sounding string.

As an aside, this is a much more elegant analogy of the other significant 
oscillation phenomenon in particle physics, that of neutral kaons. These 
are made as  flavour eigenstates $K^0$ and $\bar K^0$, but these are not 
mass 
eigenstates (just like neutrinos). Whichever one you make, it propagates 
as mix ($K^0+\bar 
K^0$ or $K^0-\bar
K^0$) of two states which have radically different decay times (called 
``K-long" and ``K-short"). One can even, in the string analogy, regenerate 
the 
rapidly decaying 
component by damping one string, in the same manner as one regenerates K-short
particles by placing a thin piece of material in the beam, thus 
preferentially ``damping-out" the $K^0$s.

\begin{figure}
\includegraphics[width=10cm]{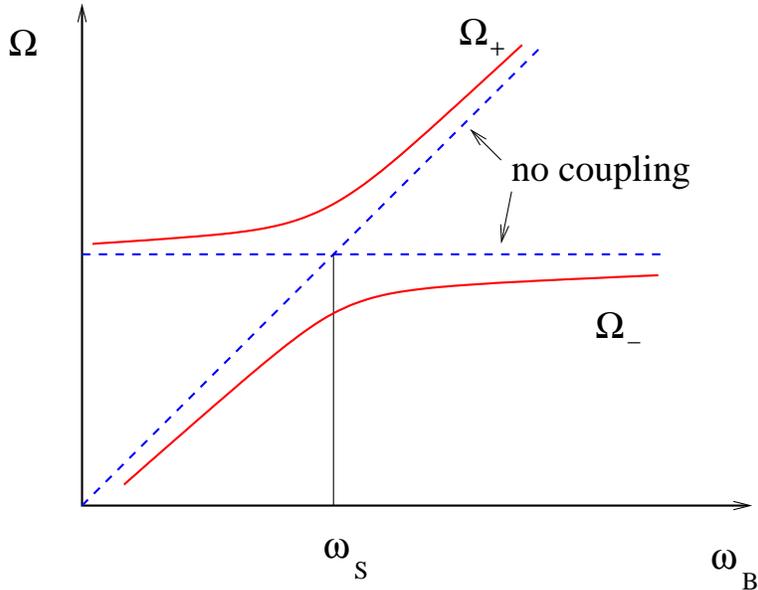}
\caption{Resonant frequencies $\Omega_\pm$ of a system consisting of 
a sound board with a 
characteristic frequency
$\omega_B$ and two identical strings with frequency $\omega_S$.
The dashed line shows the case where there is no coupling between 
string and sound board.
}
\label{fig:level} \end{figure}

However, if one now considers a sound board to have a characteristic 
frequency
$\omega_B$ and two identical string with frequency $\omega_S$, then one 
finds that the resonant frequencies of the whole system $\Omega_\pm$ 
behave as 
shown in fig.\ref{fig:level}\cite{gough}.
  
With no coupling one has two frequencies $\omega_B$ and $\omega_S$. With 
coupling the same is true so long as $\omega_B$ is either much higher or 
much lower than $\omega_S$. The interesting part occurs when the two 
frequencies are very similar, and this is an analogy with what happens 
when $\nu_2$ are born and travel out from the solar core. At birth in the 
dense core, $m_2$ is dictated by fundamental physics and the 
electron density (i.e. the sound board frequency ``$\omega_B$"). As it 
leaves the Sun, this state passes through the resonance region and emerges 
as the higher of the two mass states (``$\Omega_+$"), the one with 25\% 
electron content.

Incidentally, the SMA solution involved a rapid (``non-adiabatic") passage 
through the  
resonance region which allowed a hop from the $\Omega_+$ to $\Omega_-$
states, even with a tiny coupling. As we have seen, however, nature did 
not go this route.

\section{Conclusion}

In neutrinos we are confronted by real particles which behave in a quantum
mechanical fashion over very large distances, 1000s of km. We are not used
to this, and the physics community has yet to assimilate fully the
implications.

\section*{Acknowledgments}

Many thanks to my 
colleagues, students and seminar audiences  for reading and commenting 
on this material. Particular thanks to Douglas Scott for prodding me into 
writing it up and using a provocative title.

\end{document}